\documentstyle[psfig]{mn2e}
\input epsf

\def\lesssim{\mathrel{\hbox{\rlap{\hbox{\lower4pt\hbox{$\sim$}}}\hbox{$<$}}}}
\def\gtrsim{\mathrel{\hbox{\rlap{\hbox{\lower4pt\hbox{$\sim$}}}\hbox{$>$}}}}

\newcommand{\msun}{\mbox{$M_{\odot}$}}

\newcommand{\lsun}{\mbox{$L_{\odot}$}}
\newcommand{\Lsun}{\mbox{$L_{\odot}$}}

\begin{document}

\title[Young stars towards Cyg OB2 and DR 15]
{IPHAS discoveries of young stars towards Cyg OB2 and its southern periphery} 
\author[Jorick S. Vink et al. ]
{Jorick S. Vink$^{1,2}$, Janet E. Drew$^{2,3}$, Danny Steeghs$^{4,5}$, Nick J. Wright$^6$, 
\newauthor Eduardo L. Martin$^{7,8}$, Boris T. G\"ansicke$^4$, Robert Greimel$^{9,10}$, Jeremy Drake$^{5}$\\ 
$^1$Armagh Observatory, College Hill, Armagh BT61 9DG, Northern Ireland, U.K.\\
$^2$Imperial College of Science, Technology and Medicine,
    Blackett Laboratory, Exhibition Road, London,  SW7 2AZ, U.K.\\
$^3$Centre for Astrophysics Research, University of Hertfordshire, College Lane, Hatfield AL10 9AB, U.K.\\
$^4$Department of Physics, University of Warwick, Coventry, CV4 7AL, U.K.\\ 
$^5$Harvard-Smithsonian Center for Astrophysics, 60 Garden Street, 
Cambridge, MA 02138, U.S.A.\\
$^6$University College London, Department of Physics \& Astronomy, 
Gower Street, London WC1E 6BT, U.K.\\
$^7$Instituto de Astrofisica de Canarias, 38200 La Laguna, Tenerife, Spain\\
$^8$Physics Department, University of Central Florida, Orlando, FL 32816, U.S.A.\\
$^9$Isaac Newton Group of Telescopes, Apartado de correos 321,
    E-38700 Santa Cruz de la Palma, Tenerife, Spain\\
$^{10}$Institut f\"ur Physik, Karl-Franzen Universit\"at Graz,
Universit\"atsplatz 5, 8010 Graz, Austria\\
}

\date{received,  accepted}

\maketitle

\begin{abstract}
We report on the discovery of over 50 strong H$\alpha$ emitting objects towards the 
large OB association Cyg OB2 and the H{\sc ii} region DR 15 on its southern periphery. 
This was achieved using the INT Photometric H$\alpha$ 
Survey of the Northern Galactic Plane (IPHAS), combined with follow-up spectroscopy using
the MMT multi-object spectrometer HectoSpec. We present optical spectra, supplemented with 
optical $r'$, $i'$ and $H\alpha$ photometry from IPHAS, and near-infrared $J$, $H$, and $K$ 
photometry from 2MASS.
The position of the objects in the $(J - H)$ versus $(H - K)$ diagram strongly suggests 
most of them are young. Many show Ca {\sc ii} IR triplet emission 
indicating that they are in a pre-main sequence phase of evolution of T Tauri and Herbig 
Ae nature. 
Among these, we have uncovered pronounced clustering of T~Tauri stars roughly a degree south of the 
centre of Cyg OB2, in an arc close to the H{\sc ii} region DR 15, and the
radio ring nebula G79.29$+$0.46, for which we discuss its candidacy
as a luminous blue variable (LBV).   
The emission line objects toward Cyg OB2
itself could be the brightest most prominent component of a population of 
lower mass pre-main sequence stars that has yet to be uncovered.
Finally, we discuss the nature of the ongoing star formation in Cyg OB2 and
the possibility that the central OB stars have triggered star formation in the periphery. 
\end{abstract}


\section{Introduction}
\label{s_intro}

Cygnus OB2, with an age in the range of 1 to 4 Myrs has been proposed to
consist of approximately 2600 OB stars, including 120 O stars (Kn\"odlseder 
2000).  With a total mass up to $10^{5}$ \msun, Kn\"odlseder (2000) 
argued it might be a young globular cluster in the plane of the 
Milky Way, although both its relatively large spatial scale (some 30 pc), 
and its mass (Hanson 2003), are likely to prevent it from being a true 
Galactic analogue of the super-star clusters (SSCs) found in extragalactic 
systems such as M82 (Smith \& Gallagher 2001) and M51 (Bastian et al. 2005). 
Nonetheless with a mass of order $10^4$ $\msun$ (e.g. Hanson 2003), 
it offers, along with the Arches and Quintuplet clusters in the Galactic centre 
(e.g. Figer, McLean \& Morris 1999), NGC 3603 (Stolte et al. 2006), 
and the massive cluster Westerlund 1 (Clark et al. 2005), the best insight 
into the dominant mode of (massive) star formation in the universe.  It 
is undoubtedly the most massive OB association accessible from the northern
hemisphere, and has been discussed in the literature since the seminal
study by Reddish et al. (1966).  As a key feature within the highly
complex Cygnus-X region (Wendker 1984, Odenwald \& Schwartz 1993), it is
noteworthy that its extent and distance remain controversial (cf.
Comeron et al. 2002 versus Hanson 2003). 

At a distance of not less than 1.2 kpc (Hanson 2003) and 
probably no further than $\sim$ 1.7 kpc (e.g. Torres-Dodgen 
et al. 1991), Cyg OB2 offers the opportunity for optical studies of star 
formation as a function of stellar mass in a very massive cluster.   Many 
years ago, one of the 
differences between low- and high-mass star-formation was believed to be 
that the two processes occurred in very different regions. T associations, 
such as Taurus, were thought to be the dominant birth sites for low-mass 
(T Tauri) stars, while OB associations were supposed to be the nurseries 
for massive stars. This picture is no longer widely held.  Many low-mass young 
stars are found in exactly the same regions as the massive OB stars 
(e.g. Pozzo et al. 2003, Arias et al. 2007), and observations of the proplyds in 
Orion (O'Dell et al. 1993) suggest that 
the UV radiation from massive stars may evaporate the accretion disks 
around the low-mass stars, which may modify the initial mass 
function (IMF; e.g. Robberto et al. 2004).

 
To date, most PMS stars of T Tauri and Herbig type we know are found in 
relatively low total-mass low-concentration star-forming regions within 
Gould's Belt.  In comparison, we lack the full picture
of star formation in denser, more extreme environments within our own
Galaxy where, nevertheless, the opportunity to probe deeply and at fine 
spatial scales is opening up.  Finding these PMS stars, via their tell-tale 
H$\alpha$ emission, in a diverse range of 
clusters across the galactic plane is one of the opportunities provided 
by the INT Photometric H$\alpha$ Survey of the Northern Galactic Plane: 
the IPHAS survey (Drew et al. 2005). 

Here we present spectroscopic follow-up 
of candidate H$\alpha$-emitting point sources in the centre and to the
south of Cyg OB2.  In so doing, we report the first discoveries of 
T Tauri and Herbig Ae stars in such a massive OB association.  We also
identify what are most plausibly young stars that have formed in
association with DR 15, the H{\sc ii} region to the south, first picked out at
radio wavelengths by Downes \& Rinehart (1966).  The distance to DR 15
is commonly quoted as $\sim$1 kpc (e.g. Wendker et al 1991). In Sect.~\ref{s_obs}, we 
describe the spectroscopic observations, and summarise the 
photometric IPHAS observations that prompted them. The spectroscopic, 
optical and near-infrared (NIR) photometric data are presented in 
Sect.~\ref{s_res}, followed by a discussion in Sect.~\ref{s_disc}.  

\begin{figure*}
\mbox{
\epsfxsize=0.7\textwidth\epsfbox{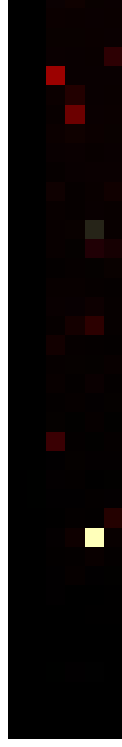} }
\mbox{
\epsfxsize=0.7\textwidth\epsfbox{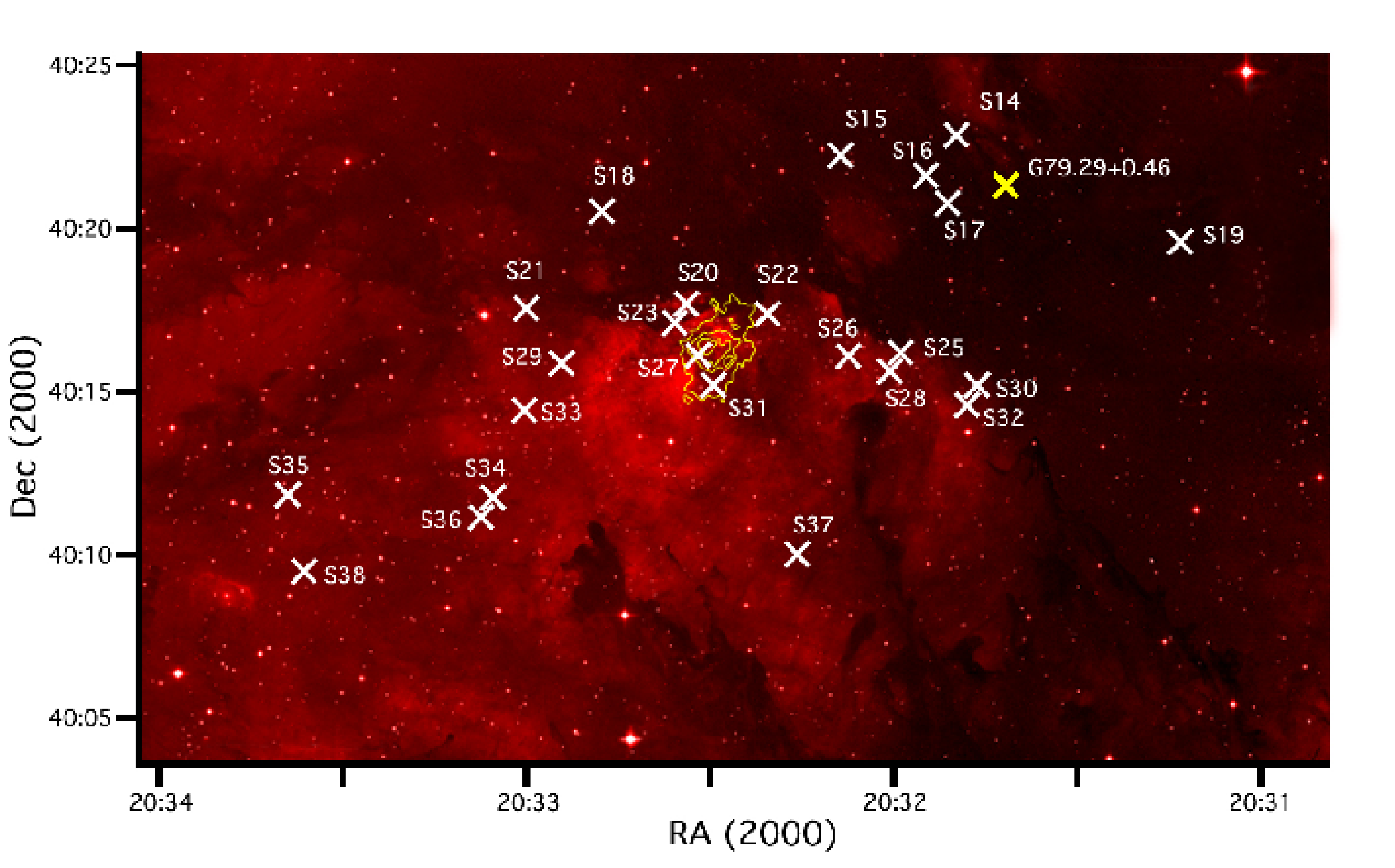}
}
\caption{A merged IPHAS H$\alpha$ image of the sky around 
the centre of Cyg OB2 (the position of the trapezium of bright O stars
Cyg OB2 No.8, as indicated with a larger cross). 
The field of view includes the locations of the confirmed line-emitting objects 
from HectoSpec pointings C (upper circle) and S (lower circle).  
The (few) objects from pointings Ea, Eb and Ec are not plotted as they fall beyond the eastern 
boundary of the image. North is up, and east is to the left. 
The lower figure is an enlarged area that is represented by the rectangular box in the upper figure.
Note that the positions of the H {\sc ii} region DR 15 and the luminous blue variable (LBV) candidate 
G79.29$+$0.46 are also indicated with contours and a cross respectively.}
\label{f_geo}
\end{figure*}

\section{Observations}
\label{s_obs}

\subsection{Spectroscopy: MMT/HECTOSPEC}

The spectra presented in this paper were obtained as part of a HectoSpec 
programme of observations aimed at giving an in-depth characterisation of 
the ($r' - H\alpha$, $r'-i'$) plane that is defined by the IPHAS survey
(Steeghs et al, in preparation, see also Drew et al 2005).  In this 
programme high priority was given to selecting relative outliers in the 
IPHAS colour-colour plane, with the result that nearly all high probability 
candidate emission 
line objects within the sampled areas, in the magnitude range  $17 < r' < 
20$, have now been followed up spectroscopically.  We refer the reader to 
section~6 in Drew et al. (2005) for more details on the target selection 
procedure. 

Spectra were obtained in either June 2004 or July 2005 using the 
multi-object spectrograph HectoSpec mounted on the Mount Hopkins 6.5-metre MMT
telescope in the F/5 configuration.  HectoSpec offers 300 fibres across a 
1-degree diameter field (Fabricant et al. 2005).   We used the 270 groove/mm 
grating, which yields a wavelength range from $\sim$4500 -- 9100 \AA\ at a 
resolution of 6.2~\AA.  The total on-source exposure times were 1200 seconds
(June 2004) and 2400 seconds (July 2005). 

The spectra were extracted using baseline products from the HectoSpec
instrument pipeline (Fabricant et al. 2005) in combination with
customised sky subtraction and flux calibration tailored for our
needs (Steeghs et al. in prep).
For each target, background sky information is provided by a collection of sky fibers
across the field of view as well as a nearby sky measurement with the same fiber
through the use of an offset exposure.
Corrections for
telluric absorption have been made, and a rough relative flux calibration 
has been applied to the spectra of targets in two of the five fields observed 
(fields C and S, see below).   In the June 
2004 data, the sky subtraction was not always perfect because of rapid
spatial variations in the diffuse and night-sky background across the field 
of view (see also Herbig \& Dahm 2001).  Hence all seeming H$\alpha$ emitters
were checked carefully in order to eliminate marginal cases where incomplete
sky subtraction may have created false positives.  
The longer July 2005 observations (in field S), 
are of higher quality and hence permit a lower acceptance threshold in 
terms of H$\alpha$ net emission equivalent width.

\begin{table}
\caption{Equatorial and galactic coordinates of the HectoSpec Pointings.}
\begin{tabular}{ccccc}
\hline
Field    & RA (2000)  & Dec (2000)  &  $\ell$ (2000) & b (2000) \\
\hline
C         &20:32:25   & $+$41:27:41 & 80.25          & $+$1.00  \\
S         &20:32:12   & $+$40:33:40 & 79.50          & $+$0.50  \\
Ea        &20:36:29   & $+$41:53:36 & 81.05          & $+$0.65  \\
Eb        &20:40:13   & $+$41:29:15 & 81.15          & $-$0.15  \\
Ec        &20:46:30   & $+$41:28:50 & 81.87          & $-$1.08  \\ 
\hline
\label{t_coor}
\end{tabular}
\end{table}

Here we present spectra from the five separate pointings observed that we 
label according to their location with respect to the notional centre of the 
large OB association Cyg OB2 -- the position of the trapezium of bright O stars
Cyg OB2 No.8.  These pointings are respectively: C, overlapping the centre;
S, immediately to the south; and Ea, Eb and Ec, in order of increasing RA, 
stepping away from the association (Table~\ref{t_coor}).  The region is 
known to be subject to significant and locally-variable extinction -- in 
Cyg OB2 itself it ranges from $A_V$ of about 3 up to over $\sim$10 (Massey \& 
Thompson 1991, Hanson 2003, Albacete Colombo et al. 2007).  

The HectoSpec observations of fields C and S have resulted in the discovery of 
respectively 10 and 38 faint ($r' > 17$) line-emitting objects.  To be 
directly associated with Cyg OB2 these will need to be at distances exceeding 
1~kpc.  In the outlying Ea field and more separated Eb and Ec fields, only a 
handful of analogous objects is found.  We shall show below that the majority of these emission line stars are 
most likely pre-main sequence objects.  The locations of the major groupings 
in fields C and S are marked on the image of the Cyg OB2 environment, derived 
from the IPHAS database, that is shown in Fig.~\ref{f_geo}.  
The H$\alpha$ emitters are found both scattered around the centre of the OB 
association and in a striking cluster of $\sim$25 objects about a degree to 
the south (in the southern portion of field S), near the position of 
DR 15. 

\subsection{IPHAS photometry}

The photometric $r'$, $i'$ and $H\alpha$ magnitudes presented here are all 
taken from the IPHAS survey (Drew et al. 2005, Gonzales-Solares et al. 2008).  
Owing to reduced and variable transmission at the first attempt to image the Cyg OB2 area in
the summer-autumn of 2003, repeat observations were made during the
moonless and photometric nights of Aug 9 and 10, 2004.  These later, superior 
data form the basis for the discussion of photometry here, and they  
were also employed as the basis for the selection of HectoSpec targets 
(field S) for the July 2005 MMT run.  For fields C, Ea, Eb and Ec the 
earlier 2003 data were the only option at the time the target selections 
were made. 

We can also utilise the repeat observations to check for
emission-line variability of objects in this part of the sky. 
One object of particular interest is G79.29$+$0.46,
a candidate luminous blue variable (LBV). We comment on this
object and its evolutionary status in Sect.~\ref{s_g79}.

\section{Results} 
\label{s_res}

\begin{figure}
\mbox{
\epsfxsize=0.35\textwidth\epsfbox{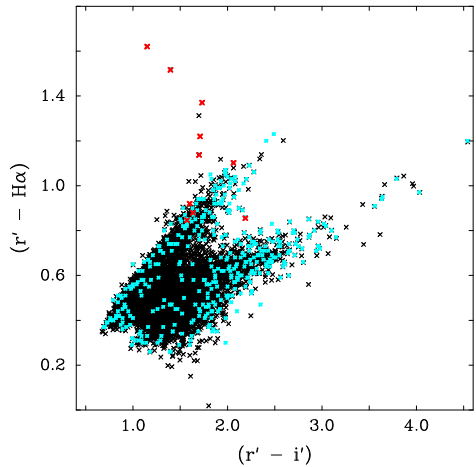}   }
\caption{The colour-colour diagram used for the selection of HectoSpec
targets in field C.  The dense locus of points plotted in black shows
the IPHAS colours of all catalogued point sources, in the magnitude range
$17 < r' < 20$,  within 30 arcmin of the field centre.  The blue data 
points are the colours of the objects for which we have HectoSpec data.
In red, we pick out the objects with H$\alpha$ emission that are listed
in Table~\ref{t_fieldC}. }
\label{f_iphasC}
\end{figure}

\begin{figure}
\mbox{
\epsfxsize=0.48\textwidth\epsfbox{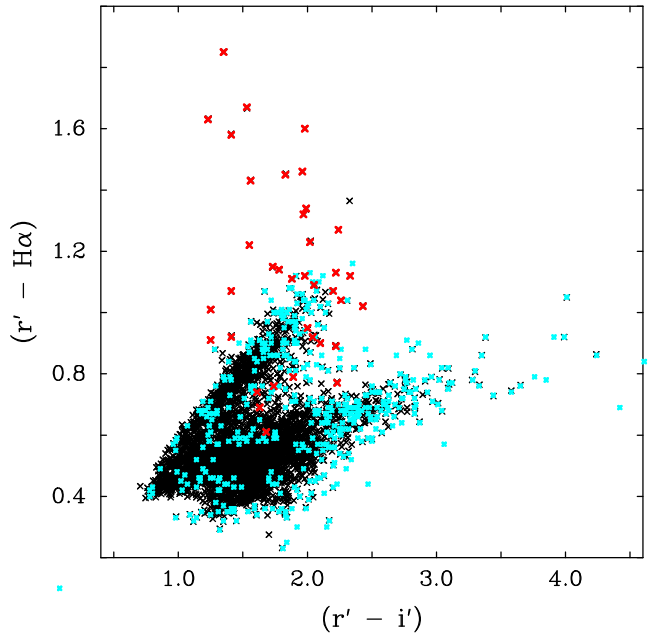}   }
\caption{The colour-colour diagram used for the selection of HectoSpec
targets in field S.  The dense locus of points plotted in black shows
the IPHAS colours of all catalogued point sources, in the magnitude range
$17 < r' < 20$,  within 30 arcmin of the field centre.  The blue data 
points are the colours of the objects for which we have HectoSpec data.
In red, we pick out the objects with H$\alpha$ emission that are listed
in Table~\ref{t_fieldS}. }
\label{f_iphasS}
\end{figure}

The emission-line objects we have found in fields C and S are listed in 
Tables~\ref{t_fieldC} and~\ref{t_fieldS} and are plotted in the IPHAS 
colour-colour plane, shown in Figs.~\ref{f_iphasC} and~\ref{f_iphasS}. 
The additional six emission-line objects from Fields Ea, Eb and Ec 
are listed in Table~\ref{t_fieldsE}.
The tables give the $r'$ magnitudes and $(r'-i')$,$(r'-H\alpha)$ colours, 
derived from the better quality IPHAS imaging obtained in August 2004. 
We also include $J$, $H$, and $K$ magnitudes from the 2MASS database, as 
these assist in constraining the nature of the line-emitting objects 
via their broad-band infrared (IR) colours. After presenting the data for 
fields C and S, we discuss the clustering in the southern part of Field S in 
close proximity to DR 15 (Downes \& Rinehart 1966), and on the LBV 
candidate G79, before briefly commenting on the 
six additional emission line stars found in the eastern fields Ea, Eb, and Ec.

\begin{figure}
\mbox{
\epsfxsize=0.5\textwidth\epsfbox{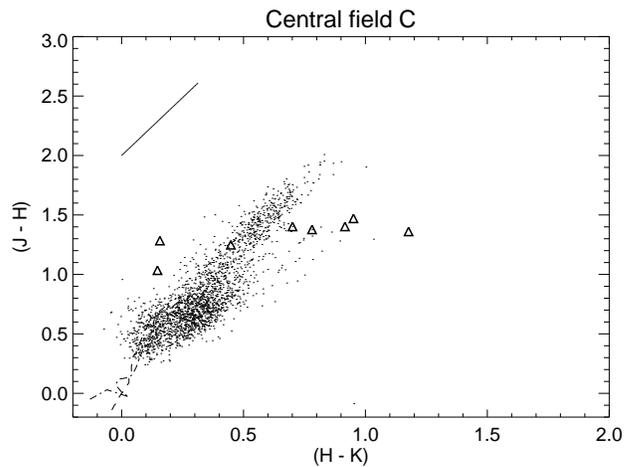}
}
\caption{2MASS data for IPHAS field 5985 (with $12 < J < 15$) and the 
strong H$\alpha$ emitting objects in the central Cyg OB2 field C. 
The solid line in the top right corner indicates the interstellar reddening line due to 
Bessell \& Brett (1988). The dashed and dashed-dotted lines in the bottom 
right corner denote the main sequence for luminosity class {\sc v} and 
{\sc iii} objects.}
\label{f_2massC}
\end{figure}

\begin{figure}
\mbox{
\epsfxsize=0.5\textwidth\epsfbox{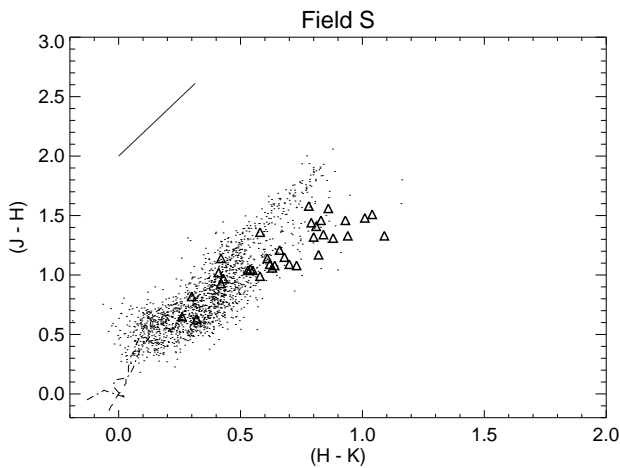}
}
\caption{2MASS data for IPHAS field 6010 (with $12 < J < 15$) and the strong 
H$\alpha$ emitting objects in Cyg OB2 field S.  The solid line in the top 
right corner indicates the interstellar reddening line due to Bessell \& 
Brett (1988). The dashed and dashed-dotted lines in the bottom right corner 
denote the main sequence for luminosity class {\sc v} and {\sc iii} objects.}
\label{f_2massG}
\end{figure}

\subsection{The centre of Cyg OB2: field C}

\subsubsection{Optical IPHAS photometry}

Figure~\ref{f_iphasC} presents the IPHAS colour-colour diagram for field C,
with the 10 H$\alpha$ emitting objects picked out in red. 
Note that sources C8 and C10 lie at the top of the IPHAS
colour-colour diagram, as expected on the basis of their large
measured H$\alpha$ equivalent widths (EW) on either side of 200\AA (250 and 180\AA,
respectively, Table~\ref{t_fieldC}).  
Similarly, the emission line objects with the smallest H$\alpha$ EW 
(e.g. C1 and C9) tend to present lower $(r'-H\alpha)$.  
Nevertheless, there is not a perfect correlation, since any of stellar 
H$\alpha$ variability, differences in the underlying spectral energy 
distribution (SED), and residual uncertainties in the sky-subtraction can 
obscure. 

We further note that 6 of the 10 emission line objects have 
$1.57 < (r'-i') < 1.73$ (C1, C2, C3, C4, C5, C7).  
C6 and C9 are even redder.   If the intrinsic colours of these objects were 
to be like those of A and earlier-type stars, their reddenings are likely to 
be $A_V \sim 7$ or more.  This is easily within the range already known to be 
typical of Cyg OB2.  Given the apparent brightness of these stars ($18.4 < r'
< 20.3$), absolute magnitudes, $M(r')$ of $\sim$2, or brighter are implied for 
a distance modulus of 11.  

On photometric grounds alone, we might conclude that these objects are 
classical Be stars within Cyg OB2, but when more evidence is taken into
account, it is more plausible they are mainly in the PMS phase, as Herbig 
or T Tauri stars. 

\begin{table*}
\caption{Positions and magnitudes for the emission line stars in the central 
field C. The $r'$ magnitudes and $(r'-i')$, $(r' - H\alpha)$ colours have been taken 
from the best IPHAS imaging (obtained in Aug 2004).  The 
H$\alpha$ EWs quoted in the final column are indicative only ($\pm$10\% ), 
as the pipeline extracted/flux-calibrated spectra are subject to zero-point 
uncertainties that particularly affect sources with faint continua. 
Note that the EW sign convention is reversed, i.e. a positive EW value 
means net emission.}
\label{t_fieldC}
\begin{tabular}{lcccccccr}
\hline
   & IPHAS name/position & \multicolumn{3}{c}{IPHAS photometry} 
   & \multicolumn{3}{c}{2MASS magnitudes} & H$\alpha$(EW)\\ 
   & J[RA(2000)$+$Dec(2000)] & $r'$ & $r'-i'$ & $r'-H\alpha$ & $J$ & $H$ & $K$ & (\AA ) \\
\hline  
C1  & J203228.09$+$414008.0 & 19.34$\pm$0.02 & 1.64$\pm$0.02 & 0.88$\pm$0.04 & 14.66$\pm$0.04 & 13.25$\pm$0.03 & 12.55$\pm$0.03 &   30\\
C2  & J203432.58$+$413641.5 & 20.09$\pm$0.03 & 1.71$\pm$0.03 & 1.22$\pm$0.05 & 15.91$\pm$0.07 & 14.63$\pm$0.05 & 14.47$\pm$0.10 &  100\\
C3  & J203258.80$+$413209.6 & 19.12$\pm$0.02 & 1.70$\pm$0.02 & 1.14$\pm$0.03 &	 14.28$\pm$0.03 & 12.90$\pm$0.03 & 12.12$\pm$0.02 &   70\\
C4  & J203200.95$+$413114.0 & 19.70$\pm$0.02 & 1.60$\pm$0.02 & 0.92$\pm$0.03 &	 15.35$\pm$0.05 & 14.10$\pm$0.04 & 13.66$\pm$0.05 &   70\\ 
C5  & J203105.70$+$412834.7 & 20.30$\pm$0.04 & 1.73$\pm$0.05 & 1.37$\pm$0.06 &	 15.16$\pm$0.05 & 13.69$\pm$0.030 & 12.74$\pm$0.03 &  105\\
C6  & J203113.63$+$412744.1 & 18.40$\pm$0.01 & 2.06$\pm$0.01 & 1.10$\pm$0.02 &	 12.75$\pm$0.02 & 11.40$\pm$0.01 & 10.21$\pm$0.02 &    60\\ 
C7  & J203224.94$+$412521.0 & 19.77$\pm$0.03 & 1.57$\pm$0.02 & 0.85$\pm$0.04 &	 15.31$\pm$0.04 & 13.91$\pm$0.03 & 13.00$\pm$0.03 &    50\\
C8  & J203219.50$+$412337.6 & 21.05$\pm$0.07 & 1.15$\pm$0.05 & 1.62$\pm$0.06 &	 $>$16.23 & 15.71$\pm$0.13 & $>$14.36 &  250\\
C9  & J203245.99$+$411042.6 & 19.72$\pm$0.03 & 2.19$\pm$0.04 & 0.86$\pm$0.04 &  13.53$\pm$ 0.04 & 12.00$\pm$0.04 & $>$11.00    &    15\\
C10 & J203217.27$+$410225.6 & 19.31$\pm$0.02 & 1.40$\pm$0.03 & 1.52$\pm$0.03 &	 15.56$\pm$0.06 & 14.53$\pm$0.07 & 14.38$\pm$0.10 &   180\\   
\hline
\end{tabular}
\end{table*}

\subsubsection{Near infrared 2MASS photometry}

In order to better pin down the character of the H$\alpha$ emitting objects, 
and to test for youth via the presence of accretion disks, 
we have examined NIR photometry from the 2MASS all-sky survey. 
A strong dusty NIR excess would distinguish the line-emitting objects 
from classical Be stars which fall in a different regime within the 
$(H - K)$ versus $(J - H)$ colour-colour diagram (e.g. see Figure~2 in 
Corradi et al. 2008). 

Eight of the emission line objects within field C are plotted in the 2MASS 
$(J-H)$ versus $(H-K)$ colour-colour diagram in Fig.~\ref{f_2massC}, along
with point sources from IPHAS field 5985 (lying inside field C).  
Objects C5, C6, C7 and (marginally) C3 are found to the right-hand side 
of the main locus of field stars and below the OB-star reddening line, 
indicating an NIR excess. Such an excess can be a signature of a 
circumstellar disk around a young pre-main sequence T Tauri or Herbig 
Ae/Be star, although it only implies the {\it presence} of dust and it does 
not provide information on the circumstellar geometry, as NIR excesses may 
sometimes be due to (or confused with) infrared companion stars (see e.g. 
Duchene et al. 2003 on V773 Tau). 

Objects C4, C10, C2 and (marginally) C1 exhibit NIR colours associated with 
intrinsically red late-type objects within or above the reddened main 
sequence band.  C8 and C9 cannot be plotted in Fig.~\ref{f_2massC} because one 
or more of their $JHK$ magnitudes is an upper limit.

\subsubsection{Optical spectroscopy with MMT/HectoSpec}

\begin{figure}
\mbox{
\epsfxsize=0.50\textwidth\epsfbox{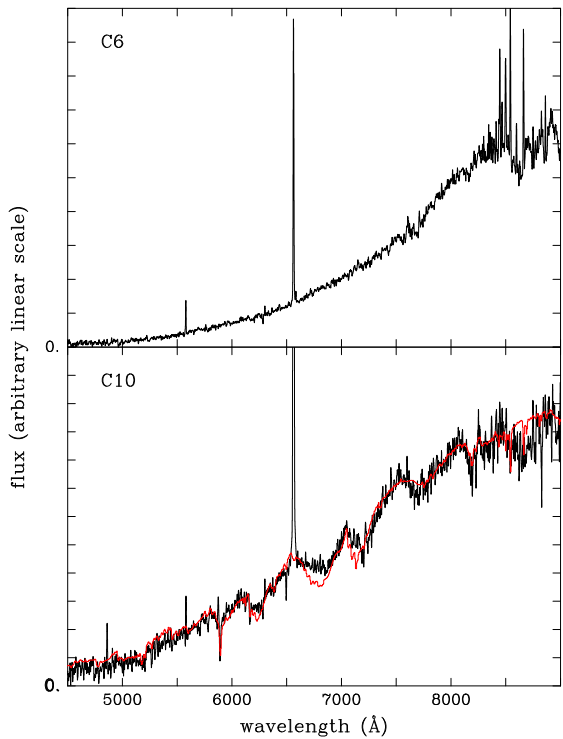} }
\caption{The spectra of C6 and C10.  The data have been 
approximately flux calibrated, and the vertical scales are linear 
in both panels. C6, in the upper panel, is representative of 5-6 objects 
from field C in presenting a significantly reddened, featureless continuum 
(at the modest S/N of the data): it is one of 4 in field C to show the 
Ca{\sc ii} IR triplet in emission. Object C10, on the other hand, shows clearly 
the molecular bands typical of an M dwarf.  Superimposed in red is the
spectrum of an M2V star (Pickles 1998) reddened by 1.9 magnitudes.
Objects C1 and C2 and C4 may be similar.}
\label{f_C_spectra}
\end{figure}

The final characterisation of the discovered emission-line stars is 
essentially upheld by the optical Hectospec data, in which we have looked 
for characterising spectral indicators.  Since the work of Hamann \& Persson 
(1992a,b) it has been recognised that the Ca{\sc ii} IR triplet is a useful 
tool to identify T Tauri and Herbig Ae/Be stars, with about half of them
showing it in emission.  We show the best-exposed representative examples, 
C6 and C10, in Fig.~\ref{f_C_spectra}.  

The NIR-excess objects, C3, C5 and C6 show the Ca{\sc ii} IR triplet in 
emission, with only C3 showing a hint of M-star molecular bands.  The
non-excess objects C8 and C9 also show the triplet in emission, and share the
apparently featureless continua of C5 and C6.  C7 is the only NIR-excess 
object in which the Ca~{\sc ii} lines are not detected in emission: at the
poor S/N achieved, its continuum is also apparently featureless.

In all of C1, C2, C4, and C10, there is evidence of the broad 
molecular bands typical of M stars.  In the best exposed example, C10, 
these features are seen at greatest contrast.  In this case, the spectrum 
can be dereddened, assuming $A_V$ is $\sim2$, to fit it acceptably well to the 
M2V spectrum in the Pickles (1998) library (see Fig.~\ref{f_C_spectra}).  
The relatively low reddening needed for this, compared with the typical range 
for Cyg OB2, could indicate C10 is a foreground object.  However the 
combination of great H$\alpha$ equivalent width and M2 spectral type 
identifies it as a T Tauri star (Barrado Navascues \& Martin 2003).  Hence 
it would be plausible to find C10 in or close to the main Cyg OB2 association 
at DM $\sim$ 11.  At an $r'$ magnitude of 19.3, and for $A_V \sim 2$, a 
distance modulus of 11 implies an absolute $M(r')$ magnitude of $\sim$6.5 
for C10.  Correcting this to V for an early M star yields $V \sim 7.5$, 
which is around 3 magnitudes brighter than expected for the early-M main 
sequence (Jahreiss \& Wielen 1997). This is compatible 
with expectation for a T Tauri star (see e.g. Kenyon \& Hartmann 1995).  

No object in Table~\ref{t_fieldC} has a bluer optical continuum than C10.
Given how faint all these emission line stars are, it would be more 
contrived to place any of them in the foreground than to adopt the 
working hypothesis that the bulk of these objects are PMS stars falling within 
Cyg OB2's sphere of influence at DM $\sim$ 11.  

\subsection{To the south of Cyg OB2: Field S}

\subsubsection{The clustering linked to the DR 15 region}

From Figure~\ref{f_geo}, it is very clear that around half of all the 
confirmed emission line stars are concentrated within as little as a tenth 
of the observed sky area.  Furthermore it can be seen that the H{\sc ii} region 
DR 15 is located in the midst of this grouping.  Hence the first possibility 
to consider is that these emission line stars belong to the same star forming 
cloud as DR 15.

Balloon observations by Emerson et al. (1973) identified the H{\sc ii} region 
DR 15 (Downes \& Rinehart 1966) to be an unresolved infrared source with a 
total luminosity exceeding 20~000 $\Lsun$ for an assumed distance of 1~kpc 
(the most commonly quoted distance, due to Wendker et al. 1991).   Subsequent 
near-infrared observations by Comeron \& Torra (2001), Dutra \& Bica (2001) 
and LeDuigou \& Kn\"odlseder (2002) reveal two separate structures with 
different reddenings and probably different distances.  The centre of the 
cluster is found to be very close to both the DR 15 H{\sc ii} region and IRAS 
source 20306+4005 (Parthasarathy et al. 1992). This IRAS source is also 
very close to our Field S source S31. 

For the purpose of presenting the spectra of all the field S sources, we 
divide them into two groups: the more northerly objects in the field, S1 to 
S13, are not viewed as candidate members of the DR 15 cluster, while S14 to
S38 are treated as such.  The dividing line between these two groups is
at declination $+$40deg 24m.

\subsubsection{spectral classification of Field S objects}

Figure~\ref{f_iphasS} represents the IPHAS colour-colour diagram for the 
38 line-emitting objects found in Field S. The photometry and
measured H$\alpha$ equivalent widths are set out in Table~\ref{t_fieldS}. 
Again, to better characterise the H$\alpha$ emitting objects and to test for 
youth, we examined NIR photometry from 2MASS.  Thirty-five of the 
38 emission-line objects in field S are plotted in the 2MASS $(J-H)$ versus 
$(H-K)$ colour-colour diagram in Fig.~\ref{f_2massG}, along with point 
sources from IPHAS field 6010 (overlapping field S).  The large majority of 
objects is found in or to the right-hand side of the main locus of field 
stars.  Clear NIR excesses are present in 9 of them.  

\begin{figure*}
\mbox{
\epsfxsize=0.8\textwidth\epsfbox{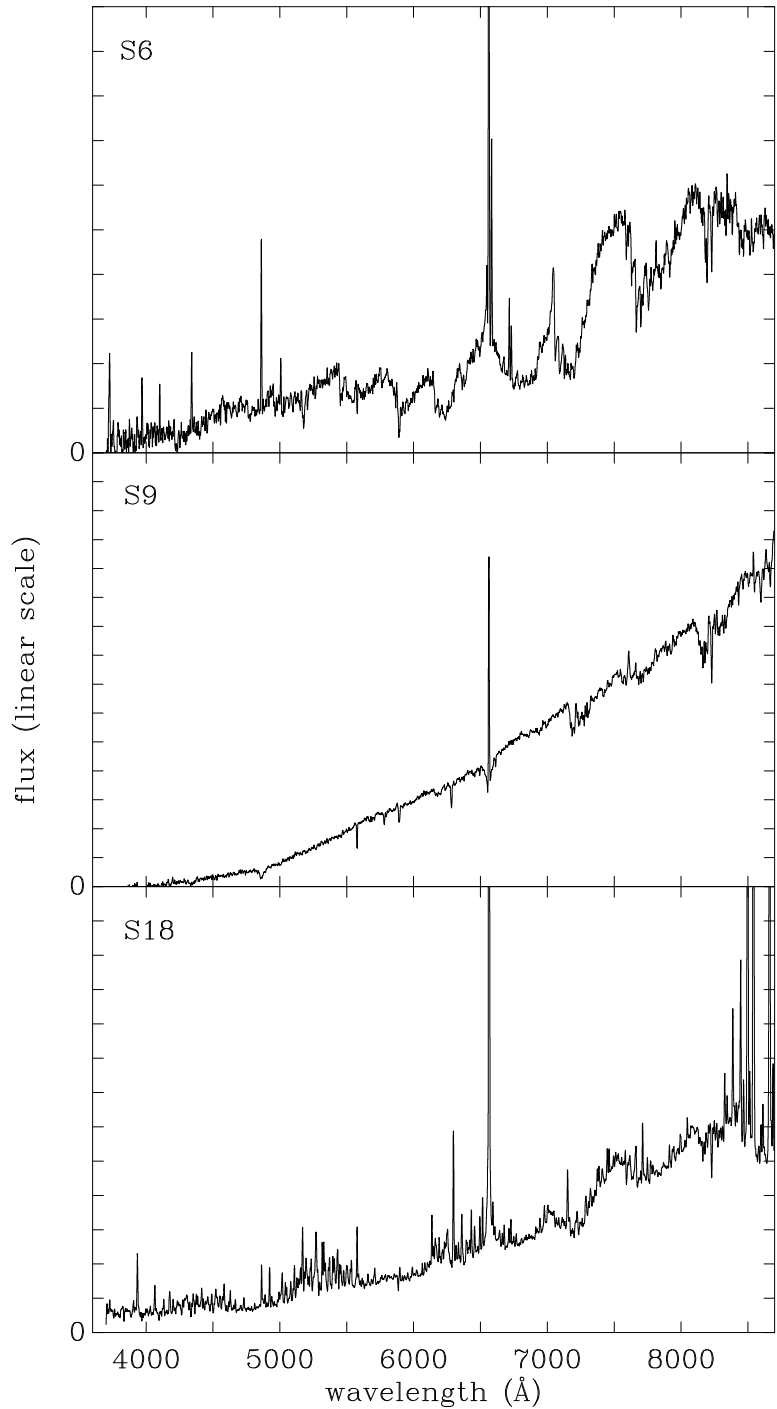}  }
\caption{The spectra of S6, S9 and S18.  The data have been 
approximately flux calibrated, and the vertical scales are linear 
in all three panels. S6 is representative of 16 objects 
from field S in showing M-type molecular bands.
S9 is more representative of the Field S emission line stars without
evident molecular bands. These are F/G/K T Tauri
stars, while the spectrum of S9 itself is more characteristic of a Herbig
Ae star.
Object S18 shows the molecular bands typical of an M dwarf, and shows the Ca{\sc ii} IR triplet 
strongly in emission. 
It has a strong H$\alpha$ line and the rich emission-line spectrum is reminiscent of that of 
DG Tau.}
\label{f_G_spectra}
\end{figure*}

The two spatial groupings do not strongly separate in terms of 
their optical line emission characteristics ($H\alpha$ EW; Ca {\sc ii} 
triplet emission).  The spectra of the 38 line-emitting objects in Field S 
were generally of higher quality than those in the central field C, allowing 
more incisive evaluation.  The Field S spectra were closely inspected and 
were divided into two spectroscopic groups: those presenting M-type molecular 
bands (Table~\ref{t_specM}), and those without them (Table~\ref{t_specG}) 
indicative of earlier spectral types. Three representative 
spectra, S6, S9 and S18 are plotted in Fig.~\ref{f_G_spectra}. 

Of the 16 objects from field S showing M-type molecular bands, object S6 has
been selected for illustration in Fig.~\ref{f_G_spectra} as it is one of the most
easily typed of our discoveries. The spectrum is a reasonable match to M4V 
(Pickles 1998) and shows little sign of reddening.  At the same time its
strong H$\alpha$ emission (EW $\sim$ 80 \AA ) clearly signals youth.  The 
type of M4 is the latest type to be assigned to any of our objects.  If we 
require S6 to be closer than 1 kpc (or 1.5~kpc), its absolute magnitude 
is fainter than $M_V \sim 8$ (or $\sim 7$).  Following the data given by
Jahreiss \& Wielen (1997), we would estimate $M_V \sim 12$ for an object 
already on the main sequence.

All objects in the M-type table satisfy the criteria based on spectral 
type and H$\alpha$ equivalent width of Barrado Navascues \& Martin (2003) 
that allow them to be designated classical T Tauri stars. 
Atomic line EWs are given in columns (2) and (3) of Table~\ref{t_specM}, 
whilst the characterising molecule indices I1,I2 and I3 of Martin \& Kun 
(1995) are given in columns (4)-(6). Note that, in some stars, the I3 index 
denominator is affected by HeI emission at 6678 \AA.   The spectral type, 
derived from indices, I1--I3, is given in column (7), whist the spectral 
type estimated from the appearance of the spectrum longward of 7000~\AA\ is 
given in column (8). Where this second type is noticeably later than that 
implied by the indices I1--I3 from column (7), it is likely that 
significant continuum veiling is also present (see notes in column 9). 
All but two of the objects in Table~\ref{t_specM} are located in the 
southern part of the field near to DR 15.

S9 is shown as an example of the Field S emission-line stars without
evident molecular bands (Table~\ref{t_specG}).   On the sky, these 
objects fall more evenly between the more northerly and DR~15 groups.
F, G and K spectral types are most likely to apply to this selection,
and many are likely to be young stars on account of their H$\alpha$
equivalent widths, and the high frequency of CaII IR-triplet emission
(around half of the list).  The spectrum of S9 could be sufficiently
early in type to be described as a Herbig Ae star. The presence of atomic 
emission (columns 2-4, in table~\ref{t_specG}) and absorption (columns 5 
\& 6) features is used, where possible, to roughly appraise spectral
type (column 7). 

Finally, we present the spectrum of object S18 (bottom panel of 
Fig~\ref{f_G_spectra}). It shows molecular bands that are typical of an M dwarf. 
The Ca {\sc ii} infrared emission lines are spectacularly strong. 
Even with the 6.2~\AA\ resolution of the HectoSpec spectra their peak
fluxes are almost ten times the continuum level.
Unusually, S18 also shows a 
very rich array of further low-excitation atomic emission lines in its 
spectrum - a property it shares with the well-known T Tauri star, DG Tau 
(see Hessman \& Guenther 1997).

\begin{table*}
\caption{Positions and magnitudes for the emission line stars in Field S.}  
\label{t_fieldS}
\begin{tabular}{lcccccccr}
\hline
   & IPHAS name/position & \multicolumn{3}{c}{IPHAS photometry} 
   & \multicolumn{3}{c}{2MASS magnitudes} & H$\alpha$ EW \\ 
   & J[RA(2000)$+$Dec(2000)] & $r'$ & $r'-i'$ & $r'-H\alpha $ & $J$ & $H$ & $K$ & (\AA ) \\
\hline 
\multicolumn{3}{l}{Objects not clustered near DR 15} \\
S1     & J203317.17$+$410015.5 & 19.85$\pm$0.03 & 2.20$\pm$0.04 & 1.07$\pm$0.05 & 13.97$\pm$0.05 & 12.51$\pm$0.04 & 11.68$\pm$0.03 & 30 \\
S2     & J203124.76$+$405944.3 & 20.30$\pm$0.04 & 1.97$\pm$0.05 & 1.32$\pm$0.06 &  14.77$\pm$0.04 & 13.26$\pm$0.03 & 12.22$\pm$0.02 & 120: \\
S3     & J203310.18$+$405903.7 & 18.16$\pm$0.01 & 1.89$\pm$0.01 & 0.79$\pm$0.02 & 13.05$\pm$0.02 & 11.72$\pm$0.02 & 10.78$\pm$0.02 & 12 \\
S4     & J203353.44$+$405449.1 & 18.78$\pm$0.01 & 1.96$\pm$0.01 & 1.46$\pm$0.02 &  13.04$\pm$0.02 & 11.70$\pm$0.02 & 10.86$\pm$0.01 & 125: \\
S5     & J203132.56$+$405138.5 & 20.34$\pm$0.05 & 1.80$\pm$0.06 & 1.25$\pm$0.07 & 16.16: & 14.58$\pm$0.06 & 14.11$\pm$0.07 &  70 \\
S6     & J203043.02$+$405033.6 & 19.62$\pm$0.02 & 1.89$\pm$0.02 & 1.24$\pm$0.04 & 15.85$\pm$0.07 & 15.22$\pm$0.09 & 14.90$\pm$0.14 & 80 \\
S7     & J203123.14$+$405024.9 & 18.00$\pm$0.01 & 1.61$\pm$0.01 & 0.74$\pm$0.02 & 13.47$\pm$0.03 & 12.32$\pm$0.03 & 11.64$\pm$0.02 &  20 \\
S8     & J203314.89$+$404909.8 & 20.37$\pm$0.05 & 2.24$\pm$0.05 & 1.27$\pm$0.07 & 14.48$\pm$0.03 & 13.07$\pm$ 0.02 & 12.26$\pm$0.02 &  70 \\
S9     & J203253.33$+$404823.7 & 17.54$\pm$0.01 & 1.68$\pm$0.01 & 0.61$\pm$0.01 & 12.82$\pm$0.02 & 11.49$\pm$0.02 & 10.40$\pm$0.02 &  20 \\
S10    & J203101.33$+$404200.7 & 19.75$\pm$0.03 & 1.73$\pm$0.04 & 1.15$\pm$0.04 &  15.00$\pm$0.05 &  $>$13.55 &  $>$12.764 & 110: \\
S11    & J203206.86$+$403952.8 & 19.56$\pm$0.02 & 2.10$\pm$0.03 & 0.90$\pm$0.04  & 14.04$\pm$0.03 & 12.72$\pm$0.03 & 11.92$\pm$0.02 &  20 \\
S12    & J203125.87$+$403231.8 & 19.58$\pm$0.03 & 2.04$\pm$0.03 & 0.92$\pm$0.04  & 14.37$\pm$0.03 & 13.23$\pm$0.03 & 12.81$\pm$0.03 &  20 \\
S13    & J203347.69$+$402547.1 & 19.07$\pm$0.02 & 2.43$\pm$0.02 & 1.02$\pm$0.03  & 12.67$\pm$0.03 & 11.50$\pm$0.02 & 10.68$\pm$0.02 &  40 \\
\hline
\multicolumn{3}{l}{Objects located in the vicinity of DR~15}\\
S14    & J203150.28$+$402333.1 & 20.32$\pm$0.05 & 1.98$\pm$0.06 & 1.12$\pm$0.07  & 14.93$\pm$0.04 & 13.57$\pm$0.03 & 12.99$\pm$0.03 &  60 \\
S15    & J203209.72$+$402253.6 & 18.91$\pm$0.01 & 1.88$\pm$0.02 & 1.11$\pm$0.02 &   --  &  --  &  --  & 60 \\ 
S16    & J203155.32$+$402216.8 & 19.96$\pm$0.03 & 1.99$\pm$0.04 & 1.34$\pm$0.04 & 14.66$\pm$0.03 &  13.22$\pm$0.03 &  12.43$\pm$0.02 & 90 \\
S17    & J203151.67$+$402128.6 & 19.89$\pm$0.03 & 2.22$\pm$0.03 & 0.89$\pm$0.05  & 14.08$\pm$0.03 & 12.52$\pm$0.02 & 11.66$\pm$0.02 &  15 \\
S18    & J203248.51$+$402105.0 & 17.69$\pm$0.01 & 1.55$\pm$0.01 & 1.22$\pm$0.01 & 13.91$\pm$0.02 & 12.83$\pm$0.02  & 12.10$\pm$0.022 & 100: \\ 
S19    & J203113.61$+$402013.0 & 20.12$\pm$0.04 & 2.26$\pm$0.04 & 1.04$\pm$0.06  &14.09$\pm$0.03 & 12.51$\pm$0.02 & 11.73$\pm$0.02 &  20 \\
S20    & J203234.88$+$401811.1 & 17.01$\pm$0.00 & 1.25$\pm$0.01 & 0.91$\pm$0.01 & 13.59$\pm$0.02 & 12.53$\pm$0.02 & 11.90$\pm$0.02 & 20 \\
S21    & J203300.81$+$401800.4 & 17.53$\pm$0.01 & 1.23$\pm$0.01 & 1.76$\pm$0.01 & 14.21$\pm$0.03 & 13.17$\pm$0.03 & 12.62$\pm$0.03 & 200: \\
S22    & J203220.50$+$401755.7 & 19.62$\pm$0.02 & 2.00$\pm$0.03 & 0.95$\pm$0.04 & 14.91$\pm$0.05 & 13.70$\pm$0.05 & 13.04$\pm$0.04 &  20 \\
S23    & J203235.81$+$401745.1 & 18.45$\pm$0.01 & 1.63$\pm$0.01 & 0.69$\pm$0.02 & 13.91: & 12.71$\pm$0.03 & 12.11$\pm$0.03 &  10 \\
S24    & J203017.09$+$401652.9 & 20.22$\pm$0.04 & 2.22$\pm$0.04 & 1.13$\pm$0.06 & 12.68$\pm$0.02 & 11.86$\pm$0.02 & 11.56$\pm$0.02 &  50 \\       
S25    & J203159.07$+$401648.2 & 18.72$\pm$0.01 & 1.83$\pm$0.02 & 1.45$\pm$0.02 & 14.48$\pm$0.03 & 13.55$\pm$0.02 & 13.13$\pm$0.03 & 250: \\
S26    & J203207.81$+$401636.5 & 20.03$\pm$0.03 & 1.98$\pm$0.04 & 1.60$\pm$0.04 & 15.19$\pm$0.04 & 14.20$\pm$0.04 & 13.62$\pm$0.05 &  120: \\
S27    & J203232.72$+$401632.0 & 17.46$\pm$0.01 & 1.41$\pm$0.01 & 1.07$\pm$0.01 &  13.82$\pm$0.02 & 12.73$\pm$0.02 & 12.11$\pm$0.02 & 45 \\
S28    & J203200.66$+$401622.1 & 18.66$\pm$0.01 & 1.78$\pm$0.01 & 1.14$\pm$0.02 &  14.37$\pm$0.03 & 13.33$\pm$0.03 & 12.80$\pm$0.03 & 60 \\
S29    & J203255.04$+$401617.4 & 17.14$\pm$0.01 & 1.25$\pm$0.01 & 1.01$\pm$0.01 &  13.47$\pm$0.02 & 12.38$\pm$0.02 & 11.68$\pm$0.02 & 40 \\
S30    & J203146.62$+$401542.8 & 18.03$\pm$0.01 & 1.41$\pm$0.01 & 0.92$\pm$0.01 & 14.35$\pm$0.03 & 13.21$\pm$0.03 & 12.60$\pm$0.03  & 20 \\
S31    & J203229.79$+$401539.8 & 18.71$\pm$0.01 & 2.05$\pm$0.01 & 1.09$\pm$0.02 & 14.22$\pm$0.04 & 13.20$\pm$0.04 & 12.79$\pm$0.04 &  28 \\
S32    & J203149.02$+$401538.1 & 19.25$\pm$0.02 & 1.53$\pm$0.02 & 1.67$\pm$0.02 &  15.30$\pm$0.04 & 14.33$\pm$0.05 & 13.90$\pm$0.06 & 170: \\
S33    & J203301.12$+$401449.4 & 19.31$\pm$0.02 & 1.35$\pm$0.03 & 1.85$\pm$0.02 &  14.98$\pm$0.05 & 13.50$\pm$0.04 & 12.49$\pm$0.02 & 450: \\
S34    & J203306.07$+$401211.9 & 20.25$\pm$0.04 & 2.33$\pm$0.04 & 1.12$\pm$0.05 & 15.27$\pm$0.05 & 14.62$\pm$0.06 & 14.36$\pm$0.09 &  60 \\
S35    & J203339.55$+$401202.7 & 19.50$\pm$0.02 & 2.23$\pm$0.03 & 0.77$\pm$0.04 & 13.60$\pm$0.02 & 12.14$\pm$0.02 & 11.21$\pm$0.02  & 15 \\
S36    & J203309.80$+$401154.8 & 18.66$\pm$0.01 & 1.56$\pm$0.02 & 1.43$\pm$0.02 & 14.72$\pm$0.04 & 13.41$\pm$0.03 & 12.53$\pm$0.02 &  140 \\
S37    & J203215.98$+$401023.6 & 19.06$\pm$0.01 & 1.74$\pm$0.02 & 0.76$\pm$0.02 & 14.80$\pm$0.04 & 13.72$\pm$0.03 & 13.08$\pm$0.03  & 13 \\
S38    & J203336.84$+$400939.0 & 18.76$\pm$0.01 & 2.02$\pm$0.02 & 1.23$\pm$0.02 &  13.92$\pm$0.02 & 12.87$\pm$0.02 & 12.33$\pm$0.02 & 50 \\
\hline
\end{tabular}
\end{table*}

\begin{table*}
\caption{Spectroscopic properties of the Field S emission line stars with 
M-type molecular bands. All these objects satisfy the criteria using
spectral type and H$\alpha$ equivalent width (Barrado Navascues \& 
Martin 2003), to be designated classical T Tauri stars.  Where the spectral
type estimated from comparing the appearance of the spectrum longward of
7000~\AA\ is noticeably later than that consistent with the indices, I1--I3, 
it is an indication of significant continuum veiling.}  
\label{t_specM}
\begin{tabular}{lclcccllll}
\hline
 & \multicolumn{2}{c}{EWs in \AA\ } & \multicolumn{3}{c}{Indices} 
& \multicolumn{2}{c}{Spectral type} & veiled? & comment \\
     &  [SII]     & LiI   & I1 & I2 & I3  & from I1,I2,I3 & $>$7000~\AA\ & & \\ 
     &  6717,6730 & 6708  & (CaH) & (CaH) & (TiO)  & & & & \\
\hline
S4   &  --,--    &   --   &  1.09 & 1.29 & 0.94 & M0.5$\pm$0.5   & ?
& & \\          
S6   &  -7,-7    &   --   &  1.43 & 1.68 & 1.48 & M5 $\pm$1      & M4
&  no & $E(B-V) \simeq 0$ \\   
S18  & -1.4,-2.0 &   --   &  1.06 & 1.27 & 1.09 & M0 $\pm$0.5    & M2
& & Ca{\sc ii} IR triplet emission\\
 &  &  &  &  &  &  &  &  & cf. DG~Tau\\      
S21  &  --,--    &   0.5  &  1.05 & 1.19 & 0.98 & K7/M0          & M0
& no &  Ca{\sc ii} IR triplet emission\\
S22  &  --,--    &   0.4: &  1.12 & 1.29 & 1.08 & M0.5 $\pm$0.5  & M3
& yes & \\
S24  &  --,--    &   --   &  1.23 & 1.48 & 1.49 & M3.5 $\pm$0.5  & M6
& yes & \\ 
S25  &  --,--    &   0.2  &  1.15 & 1.31 & 1.21 & M1   $\pm$0.5  & M3
& & Ca{\sc ii} IR triplet emission\\
S26  &  --,--    &   0.5  &  1.17 & 1.40 & 1.24 & M1.5 $\pm$0.5  & M3
&  &  Ca{\sc ii} IR triplet emission\\        
S27  &  --,-0.4  &   0.4  &  1.06 & 1.14 & 0.98 & K7   $\pm$0.5  & M1
& yes & Ca{\sc ii} IR triplet emission\\
S28  &  --,--    &   0.7  &  1.13 & 1.29 & 1.08 & M0.5 $\pm$0.5  &
M2-3 & &  Ca{\sc ii} IR triplet emission\\
S30  &  --,--    &   0.5  &  1.07 & 1.17 & 1.00 & K7   $\pm$0.5  & M2
& yes & \\
S31  &  -0.6,-0.4  & 0.5  &  1.21 & 1.38 & 1.32 & M2   $\pm$0.5  & M4
& & \\
S32  &  -0.4,-0.3  & 0.2  &  1.13 & 1.26 & 1.04 & M0.5 $\pm$0.5  & M3
& yes & Ca{\sc ii} IR triplet emission\\
S34  &  --,--    &   --   &  1.25 & 1.50 & 1.35 & M3.5 $\pm$0.5  & M4
& no & $E(B-V) \simeq 0.6$ \\
S37  &  --,--    &   0.8  &  1.10 & 1.21 & 0.96 & M0   $\pm$0.5  &
M0-1 & no & $E(B-V) \simeq 1$ \\
S38  &  --,--    &   0.4  &  1.14 & 1.26 & 1.04 & M0.5 +/- 0.5   & M2
& &  Ca{\sc ii} IR triplet emission\\ 
\hline
\end{tabular}
\end{table*}

\begin{table*}
\caption{Spectroscopic properties of the Field S emission line stars without 
evident molecular bands (non M-type).}  
\label{t_specG}
\begin{tabular}{lcccccll}
\hline
   & \multicolumn{3}{c}{Emission features} & \multicolumn{2}{c}{Absorption 
features} & Spectral type & Comment \\ 
   & HeI & OI & CaII & LiI & 6162,6495 &  \\
\hline  
S1 & no & no & no & yes? & no,yes & G/K? & \\
S2 & yes & yes & yes & no & no,no & &  \\
S3  & yes & yes & no & yes & yes,yes & &  \\
S5  &  &  &  &  &  &  & noisy \\
S7  & no & no & no & no & yes,yes & K? & \\
S8  & no & yes & yes & ? & no,no &  &  \\
S9  & yes & no & no & no & no,yes? & Ae & 
 $E(B-V) \simeq 2.3$ \\
S10 & yes & yes & yes & no & yes,yes & G/K & \\
S11 & no & yes & no  & no & no,no &  &  \\
S12 & no & no & no & no & no,yes & K? & \\
S13 & no & yes & yes & no & no,no & & $E(B-V) \sim 3.5$  \\
S14 & no &  & no  &  &   & G/K &  \\
S15 & yes &   & yes & no & yes,yes & G/K & \\ 
S16 & yes & yes & yes & no & yes,no & &  \\
S17 & no & no & no & no & yes,yes & K? & \\
S19 & no & no & no & no & yes,yes & G/K & \\
S20 & no & no & no & yes & yes,yes & K & \\
S23 & no & no & no & yes & yes,yes & G? & \\
S29 & yes & yes & yes & yes? & yes/yes & K & \\
S33 & yes & yes & yes & no &  & & \\
S35 & no & no? & yes? &  & no,yes & G/K? & \\
S36 & yes & yes & yes & no & yes,no & & \\
\hline
\end{tabular}
\end{table*}

\subsubsection{The status of the luminous blue variable candidate G79.29$+$0.46}
\label{s_g79}

The LBV candidate (LBVc) G79.29$+$0.46 was discovered as a ring-like radio source 
by Higgs et al. (1994). In a recent study examining the location of LBVs in 
the Hertzsprung-Russell diagram, Smith et al. (2004) positioned the LBV 
candidate G79 at an absolute magnitude of log$L/\lsun = 6.1$ for an assumed 
distance of 1.7kpc. Above, we discussed the possibility 
that the southern clustering of emission line objects may be associated 
with the DR 15 region rather than with Cyg OB2 itself. 
Given that -- on the sky -- G79 is extremely close to the discovered 
line-emitting objects, we may wish to revisit the distance and luminosity 
of G79. If G79 would be a PMS belonging to the same group of stars that 
appears to be associated with DR 15, its distance may need to be revised downward 
to $\sim$1kpc and its intrinsic luminosity would drop to about 
log$L/\lsun = 5.65$.  Although this may have consequences for its 
position on the S~Dor instability strip in the Hertzsprung-Russell Diagram 
(see Fig.~1 in Smith et al. 2004), its luminosity would still be in line 
with that of an evolved massive star rather than that of an intermediate 
mass PMS. In other words, we keep G79 in the list of LBV candidates.

For any evolved star to officially classify as an LBV (i.e. to drop the ``c'' from LBVc), the object
needs to be subject to significant spectral and photometric variations (of more than 1 magnitude)
on the timescale of years to decades (e.g. Humphreys \& Davidson et al. 1994).
This is a relevant issue, as the total number of Galactic LBVs is only of order 5-10.
Given that IPHAS re-visited the field of CygOB2 on a number of occasions over the period 2003-2005, this may enable us 
to obtain vital information on G79's evolutionary status, either confirming or eliminating it as 
a bona-fide LBV. 

The IPHAS $r'$ magnitude ($r' = 14.91$, measured Oct 16 2003) has not shown enough variability 
over the 2003-2005 period for G79 to be officially added to the LBV class.  
Monitoring over longer timescale is required. Potential $(r'-i')$ colour variability (note that $r'-i'$ $=$ 2.92 
on Oct 16 2003) could be caused by temperature and/or mass-loss 
variations. For the sake of completeness and future reference, we quote G79's ultraviolet 
magnitude of U = 19.744 $\pm 0.035$ (from UVEX data taken on 27 June 2006) and its near-infrared 
photometry from 2MASS with $J = 6.91 (\pm 0.02)$, $H = 5.29 (\pm 0.02)$, and $K = 4.33 (\pm 0.01)$. 

More significantly, we can report IPHAS variable line emission, as ($r'$ $-$ H$\alpha$) varied from 
1.05 on 16 Oct 2003, up to $r'$ $-$ H$\alpha$) $=$ 1.41 (on 9 Aug 2004), and returning 
to ($r'$ $-$ H$\alpha$) $=$ 1.08 on 1 Nov 2005. 
This excursion suggests the occurrence of changes in the mass-loss rate causing the 
H$\alpha$ EW variability. 
Previously, Voors et al. (2000) studied the spectrum of G79 and found the H$\alpha$ EW to be $\simeq$50\AA.  
Our line-emission variability, underlying the IPHAS ($r'$ $-$ H$\alpha$) changes, indicates line 
EW changes by several tens to hundreds of \AA\ (see figure 6 in Drew et al. 2005), corresponding
to EW changes by a factor of two or more.
This kind of H$\alpha$ EW variability is not extraordinary in comparison to bona-fide 
Galactic LBVs, such as AG~Car, whose H$\alpha$ EW varied from $\sim$50 \AA\ to $\sim$200\AA\ within  
a few years (Stahl et al. 2001). This is in line with predicted LBV mass-loss variability 
due to changes in the ionisation of Fe that drives the winds of LBVs (Vink \& de Koter 2002).

Although we do not have definitive proof that G79 is a true LBV, all available  
evidence points in this direction. This assertion strengthens even further when 
secondary indicators, such as the presence and morphological similarity of G79's circumstellar 
nebula to nebulae from confirmed LBVs such as AG Car are taken into account.

\subsection{Fields Ea, Eb and Ec}

\begin{table*}
\caption{Positions and magnitudes for the emission line stars in Fields Ea, Eb and Ec. } 
\label{t_fieldsE}
\begin{tabular}{lcccccccr}
\hline
   & IPHAS name/position & \multicolumn{3}{c}{IPHAS photometry} 
   & \multicolumn{3}{c}{2MASS magnitudes} & H$\alpha$ EW \\ 
   & J[RA(2000)$+$Dec(2000)] & $r'$ & $r'-i'$ & $r'-H\alpha $ & $J$ & $H$ & $K$ & (\AA ) \\
\hline  
Ea1   & J203839.90$+$420118.2 &   19.83$\pm$0.02 &  1.99$\pm$0.02 &  1.48$\pm$0.04 &  14.223$\pm$0.03 &  12.760$\pm$0.03 &  11.814$\pm$0.02 & 111 \\
\hline
Eb1   &  J204059.02$+$411051.9 &  18.017$\pm$0.01 &	1.76$\pm$0.01 &  1.40$\pm$0.01 &  12.927$\pm$0.02 &	11.174$\pm$0.02 &	9.749$\pm$0.01 & 210\\
Eb2   &  J204121.02$+$411721.6 &  18.952$\pm$0.01 &	1.51$\pm$0.01 &  1.60$\pm$0.01 &  15.552$\pm$0.05 &     13.868$\pm$0.04 &      12.736$\pm$0.03 & 102\\
Eb3   &  J204140.09$+$411228.1 &  17.390$\pm$0.01 &	1.64$\pm$0.01 &  1.39$\pm$0.01 &  12.818$\pm$0.02 &	11.431$\pm$0.02 &      10.462$\pm$0.01 & 89\\
\hline
Ec1   & J204733.98$+$413137.0 &   18.162$\pm$0.02 & 1.33$\pm$0.01 &   1.64$\pm$0.01 &   14.924$\pm$0.04	& 14.088$\pm$ 0.04 & 13.576$\pm$0.05 & 210\\
Ec2   & J204645.49$+$410700.0 &   18.532$\pm$0.02 & 1.05$\pm$0.02 & 1.23$\pm$0.01    &  15.614$\pm$0.05 & 14.222$\pm$0.04 & 13.030$\pm$0.03 & 59\\
\hline
\end{tabular}
\end{table*}

Pointings Ea, Eb, and Ec revealed six further strong H$\alpha$ emitting 
objects. All of these are found far above the unreddened main sequence, and 
their $(r'-i')$ colours, $\ga 1.5$, are in good agreement with the known 
range in reddenings for the environs of Cyg OB2.  As can also be noted from 
Table~\ref{t_fieldsE}, all objects in these fields 
show large NIR excesses, except for Ec1.  Although sources Eb1 and Eb3 show
the characteristic Ca {\sc ii} IR triplet strongly in emission, source Eb2 -- 
despite exhibiting a larger NIR excess than source Eb3 -- does not.  We note 
that the extreme emission line star Ec1, with an H$\alpha$ EW of 210\AA\, 
does not show a particularly large NIR excess.   Object Ec2 stands out for 
its combination of ``moderate'' H$\alpha$ EW of 59\AA, its relatively extreme 
NIR excess, and strong Ca {\sc ii} emission.  
 
\section{Discussion} 
\label{s_disc}

\subsection{The nature of the H$\alpha$ emitters}

On the basis of photometric survey data and highly-efficient fibre spectroscopy
as follow-up, we have discovered over 50 strong H$\alpha$ emitting objects 
towards the large OB association Cyg OB2 and the 
neighbouring H{\sc ii} region DR 15.  The greater concentration of our 
discoveries - nearly half of them - lie in an arc adjacent to the latter.
 
Ca {\sc ii} IR triplet emission is clearly seen in the spectra of 26 out of
54 emission line objects.  A similar fraction shows a dusty NIR excess that 
distinguishes them from classical Be stars.  In many of the M-type discoveries
the H$\alpha$ equivalent width is clearly too extreme to allow them to be 
classified simply as active main sequence objects.  Taking these properties
altogether, it seems likely that the bulk of our sample consists of objects in 
their pre-main sequence phase of evolution.

But how massive are these objects, and how do the more scattered Cyg OB2 group
and the more concentrated DR 15 group compare?  To begin answering these 
questions, we would ideally assign reasonably precise spectral sub-types to 
the objects, but the combination of limited spectral resolution ($\sim$ 6\AA), 
together with the complicating attributes of continuum veiling and NIR excesses, 
prevent this at the present time.  For the time-being we therefore adopt a qualitative 
approach based on typical magnitudes, likely distances, and reddenings.

After a review of the literature on the distance of Cyg OB2 (Torres-Dodgen 
et al 1991, Massey \& Thompson 1991, Hanson 2003), we have adopted a 
distance modulus (DM) of 11.0 towards the region.  The typical interstellar 
reddening into Cyg OB2 is commensurate with $E(B-V)$ $\simeq$ 2 
(Massey \& Thompson 1991, Albacete-Colombo et al 2007).  Using the standard 
factor of $R_V$ = 3.1 to convert $E(B-V)$ to visual extinction, we arrive at 
$A_V$ $\simeq$ 6-7 (or $A_{r'} \simeq 5$).  Taking the observed IPHAS 
magnitudes, $18 < r' < 20$, for sources in field C and the north of field S, 
together with the distance modulus $DM \simeq 11$ and the typical reddening, 
the absolute $M(r')$ magnitudes can be deduced to fall in the range 2 -- 4.  
On the main sequence, this magnitude range would correspond to late-A and F 
spectral types.  But some of our discoveries are less reddened M-type 
objects (e.g. the object C10 discussed in section 3.1.3), that cannot be so 
intrinsically bright -- even at the distance of Cyg OB2.  The impression to
emerge from this is that a mixture of Herbig and T Tauri stars have been 
uncovered.

The sources lying in the vicinity of DR 15 may be less distant (if we 
apply the reported figure of 1~kpc due to Wendker et al 1991), perhaps 
somewhat brighter ($17 < r' < 20$), and possibly less reddened on the 
whole ($E(B-V) \sim 1$ may be typical).  
The absolute magnitudes here are likely to fall in
the range $6 < M_{r'} < 9$ -- a somewhat fainter range than for the scattered 
Cyg OB2 candidate young stars in the more northern areas. 
This picture would be consistent with the appreciably 
higher fraction of DR 15 associated candidate objects presenting M-type spectra 
(around half, compared with one quarter for the more northern objects).  

Rather than objects in the foreground or unrelated to Cyg OB2, we 
expect the majority of objects to be directly associated with Cyg OB2 
and DR 15.  

\subsection{Comments on diagnosing environmental influences on the 
newly-revealed young stellar population}

We draw attention to our discovery of line-emitting objects in the central 
portion of Cyg OB2.  The presence of some disk accretors in potentially 
close proximity to massive OB stars may be considered significant, as the 
effect of UV photo-evaporation of young disks -- clearly operative on Orion's 
proplyds (O'dell et al. 1993) -- seems well established.
Nevertheless, we have revealed PMS objects towards CygOB2 that are embedded in 
circumstellar disks and 
which appear to be indistinguishable from the clustered objects towards the southern portion of the field. 
If the central OB stars have a pronounced effect on the evolution of PMS circumstellar disks, we 
might expect to witness a decreasing influence of OB stars when 
going from Field C PMS to the northern areas of Field S and to the area of clustered objects 
in the south (see the discussions in Guarcello et al. 2007, Mayne et al. 2007).
However, there is no evidence for such a trend in the proportion of NIR excesses,  
or in the emission line characteristics (e.g. presence/absence of the Ca {\sc ii} IR triplet and in the strength of the 
$H\alpha$ emission). 

One possibility that was discussed above is that the objects clustering towards the south are closer to us 
than the objects in the north, and in this case there is no particular reason to search for trends. 
This would still not explain the presence of circumstellar disks in the central regions 
of Cyg OB2 in its own right. The mere existence of these disks could imply 
that the influence of OB stars on proto-planetary disk evolution 
has been exaggerated or that additional effects may hide the photo-evaporation effect. 
Such a situation could be true if the central objects are on average more reddened. If these central PMS stars 
are still embedded in large amounts of molecular gas, this material may shield the disks from UV radiation more effectively, potentially 
compensating for the effect of their smaller separation from the UV emitters. 
Another option would be that the Field C PMS stars are younger, and the OB stars have 
simply not yet had enough time to evaporate the proto-planetary disks. 

Alternatively we might just be at the beginning of finding candidate PMS objects 
across the face of Cyg~OB2.  The reason being that we might have expected to find 
many more young stars in a population no more than a few million
years old, that is as rich in massive OB stars as this region, unless the majority 
of PMS consists of weak-lined T Tauri stars (WTTS) which are harder to pick out from 
their H$\alpha$ emission than classical T Tauri stars (CTTS). Albacete-Colombo 
et al. (2007) studied the central regions of Cyg OB2 at X-ray wavelengths and found 
a low fraction (of only 4.4\%) of disk-induced NIR excesses.
So far, the
list of candidate young stars towards the centre of Cyg OB2 is as long as it is towards the  
more modest DR 15 region.  A comparison in typical absolute magnitude of
candidates between the two regions may hint that the limiting 
$r'$ magnitude of $\sim$20 may need to be increased to $\sim$23 or more in
order to uncover the main young low mass population in the centre. 
The encouraging lesson of this study is that insight into the 
star-forming environment within Cyg OB2 can be gained via photometry alone, 
given the high proportion of IPHAS candidate emission line stars that are 
confirmed spectroscopically.  A deeper targeted optical photometric study, 
incorporating narrow-band H$\alpha$ observations that was appropriately 
paired with NIR data may go a long way to resolving these issues.

A final possibility worthy of mention is that the PMS 
we have discovered towards the centre of CygOB2 are in reality somewhat in the near
foreground of Cyg OB2. In other words, the central PMS could 
geometrically be at the same distance from the Cyg OB2 OB population as 
those we find in the southern area of Field S, near DR 15.  For this 
concept to work the distances to the two groups of stars would need to
be very similar.  This is not ruled out given the continuing controversy
over the distance to Cyg OB2 and its surroundings (e.g. placing everything
at $\sim$1.4 kpc is a currently available option -- see Hanson 2003) and a recent CO 
view of the Cygnus X region (Schneider et al. 2006).  
Note that 1 degree on the sky at a distance of 1-2 kpc corresponds to a length of
15--30 pc. The main difficulty would be to explain why reddenings across the 
face of Cyg OB2 may be higher than at the 'limb' (in this
picture) in the vicinity of DR 15. This difference would indeed suggest that 
Cyg OB2 is distinct and more distant. If however this were not a problem, then the PMS 
objects presented here might all be seen as the consequence of triggered star 
formation on the periphery of the main OB association.  If stellar winds 
and early supernovae in the centre of the association would have been 
responsible for this, the relevant timescale is easily short enough to
allow for this connection, as expansion at a sound speed of 10 ${\rm km/s}$ would
take $\sim$3 Myr to travel $\sim$30 pc -- a timescale compatible 
with the age of Cyg OB2.

\section*{Acknowledgments}

We would like to thank the MMT/HectoSpec team and the Isaac Newton Group 
for their assistance.  This paper makes use of data from the Isaac Newton 
Telescope, operated on the island of La Palma by the Isaac Newton Group in 
the Spanish Observatorio del Roque de los Muchachos of the Instituto de 
Astrofisica de Canarias.   The multi-object spectroscopic observations 
reported here were obtained at the MMT Observatory, a joint facility of the 
Smithsonian Institution and the University of Arizona.  
We acknowledge use of data products from the Two Micron All Sky Survey 
(2MASS), which is a joint project of the University of Massachusetts and 
the Infrared Processing and Analysis Center/California Institute of 
Technology (funded by the USA's National Aeronautics and Space 
Administration and National Science Foundation).  
DS acknowledges a STFC Advanced Fellowship as well as support through the
NASA guest observer program. NJW acknowledges an STFC funded student fellowship.

\end{document}